\documentclass[a4paper,twoside,runningheads]{llncs}

\usepackage[T1]{fontenc}
\usepackage{times}
\newcommand{\draftonly}[1]{#1}
\usepackage{hyperref}
\usepackage{amsmath}
\usepackage{amssymb}
\usepackage{amsthm}
\usepackage{stmaryrd}
\usepackage{laws}

\makeatletter
\newcounter{colwidth}
\newenvironment{eqncolumns}{\@ifnextchar[{\@eqncolumns}{\@@eqncolumns}}{\end{eqnarray}\end{minipage}\vspace{1ex}}
\def\@eqncolumns[#1]{\setcounter{colwidth}{100-#1}\mbox{}\vspace{-4ex}\\\begin{minipage}[t]{0.#1\textwidth}\begin{eqnarray}}
\def\@@eqncolumns{\setcounter{colwidth}{5}\mbox{}\vspace{-4ex}\\\begin{minipage}[t]{0.5\textwidth}\begin{eqnarray}}
\newcommand{\secondcolumn}{\end{eqnarray}\end{minipage}\begin{minipage}[t]{0.\thecolwidth\textwidth}\begin{eqnarray}}
\makeatother

\newcommand\numberthis{\addtocounter{equation}{1}\tag{\theequation}} 

\newcommand{\Compatible}{S}
\newcommand{\Command}{\mathcal{C}}
\newcommand{\Atomic}{\mathcal{A}}
\newcommand{\PseudoAtomic}{\mathcal{X}}
\newcommand{\Test}{\mathcal{T}}

\newcommand{\nondet}{\mathbin{\sqcup}}
\newcommand{\Nondet}{\mathop{\bigsqcup}}
\newcommand{\Seq}{\mathbin{;}}
\newcommand{\together}{\mathbin{\Cap}}
\newcommand{\meet}{\mathbin{\sqcap}}
\newcommand{\Meet}{\mathbin{\bigsqcap}}
\newcommand{\sync}{\mathbin{\otimes}}
\newcommand{\refsto}{\mathrel{\succeq}}

\newcommand{\defs}{\mathrel{\widehat{=}}}

\newcommand{\kw}[1]{\mathsf{#1}}
\newcommand{\Magic}{\bot}
\newcommand{\Nil}{\boldsymbol{\tau}}
\newcommand{\Skip}{\kw{skip}}
\newcommand{\Chaos}{\kw{chaos}}
\newcommand{\Abort}{\top}
\newcommand{\Term}{\kw{term}}

\newcommand{\Atom}[1]{\mathsf{#1}}
\newcommand{\Ata}{\Atom{a}}
\newcommand{\Atb}{\Atom{b}}
\newcommand{\Patx}{\Atom{x}}

\newcommand{\Fin}[1]{#1^{\star}}
\newcommand{\Om}[1]{#1^{\omega}}
\newcommand{\Inf}[1]{#1^{\infty}}
\newcommand{\spot}{\mathbin{.}}
\newcommand{\nat}{\mathbb{N}}
\newcommand{\union}{\mathbin{\cup}}
\newcommand{\inter}{\mathbin{\cap}}

\newcommand{\pstepd}{\pi}
\newcommand{\estepd}{\epsilon}
\newcommand{\stepd}{\alpha}
\newcommand{\cpstepd}{\boldsymbol{\pstepd}}
\newcommand{\cestepd}{\boldsymbol{\estepd}}
\newcommand{\cstepd}{\boldsymbol{\stepd}}
\newcommand{\cpstep}[1]{\mathop{\pstepd}#1}
\newcommand{\cestep}[1]{\mathop{\estepd}#1}

\newcommand{\guar}[1]{\mathop{\kw{guar}_{\pstepd}}#1}
\newcommand{\rely}[1]{\mathop{\kw{rely}}#1}
\newcommand{\universalrel}{\kw{univ}}

\newcommand{\Assert}[1]{\{#1\}}
\newcommand{\Post}[1]{[#1]}
\newcommand{\odotneut}{\eta}
\newcommand{\otimesneut}{I}
\newcommand{\atomicotimesneut}{\iota}

\makeatletter
\newcommand{\ChainRel}[1]{\crcr %\noalign{\penalty\interdisplaylinepenalty}\hspace*{-1em}
  #1~ &
  \@ifnextchar*{\@ChainRelCommment}{}}
\newcommand{\Why}[1]{\mbox{{\color{blue}\hspace*{1em}#1}}}
\def\@ChainRelCommment*[#1]{\Why{#1}
  \crcr & %\noalign{\penalty\interdisplaylinepenalty}
  }
\newcommand{\StartRef}[1]{\hspace*{-1.5em}(\ref{#1}) \refsto
  \@ifnextchar[{\@StartRefCommment}{}}
\def\@StartRefCommment[#1]{\mbox{#1}
  \crcr %\noalign{\penalty\interdisplaylinepenalty}
  }
\makeatother
\newcommand{\Equiv}{\ChainRel{\equiv}}
\newcommand{\Refsto}{\ChainRel{\refsto}}

\makeatletter
\def\@setmcodes#1#2#3{{\count0=#1 \count1=#3
  \loop \global\mathcode\count0=\count1 \ifnum \count0<#2
  \advance\count0 by1 \advance\count1 by1 \repeat}}

\DeclareSymbolFont{italic}{OT1}{\rmdefault}{m}{it}
\let\mathit\undefined
\DeclareSymbolFontAlphabet{\mathit}{italic}
\edef\@tempa{\hexnumber@\symitalic}
\@setmcodes{`A}{`Z}{"7\@tempa41}
\@setmcodes{`a}{`z}{"7\@tempa61}
\makeatother

\usepackage[colorinlistoftodos,textwidth=2cm]{todonotes}

\definecolor{Aqua}{rgb}{0,1,1}

\newcounter{hours}
\newcounter{minutes}
\newcommand{\printtime}{%
  \ifthenelse{\value{hours}<10}{0}{}\thehours:%
  \ifthenelse{\value{minutes}<10}{0}{}\theminutes}

\title{Restructuring a concurrent refinement algebra}
\author{
Ian J. Hayes\orcidID{0000-0003-3649-392X}
\and
Larissa A. Meinicke\orcidID{0000-0002-5272-820X}
\and
Naso Evangelou-Oost\orcidID{0000-0002-8313-6127}
}

\newbox{\MyDate}
\savebox{\MyDate}{\draftonly{ (\today\ \printtime)}}
\titlerunning{Restructuring a concurrent refinement algebra \usebox{\MyDate}}
\authorrunning{I. J. Hayes \and L. A. Meinicke \and N. Evalgelou-Oost \usebox{\MyDate}}
\institute{
School of Electrical Engineering and Computer Science, \\ 
The University of Queensland, Brisbane, Queensland 4072, Australia
  \draftonly{\\\vspace*{2ex} \today~\printtime}
}

\begin{document}

\maketitle

\begin{abstract}
The concurrent refinement algebra has been developed to support rely/\-guarantee reasoning about concurrent programs. 
The algebra supports atomic commands and defines parallel composition as a synchronous operation, as in Milner's SCCS. 
In order to allow specifications to be combined, the algebra also provides a weak conjunction operation, 
which is also a synchronous operation that shares many properties with parallel composition. 
The three main operations, sequential composition, parallel composition and weak conjunction, 
all respect a (weak) quantale structure over a lattice of commands.
Further structure involves combinations of pairs of these operations: 
sequential/parallel, sequential/weak conjunction and parallel/weak conjunction, 
each pair satisfying a weak interchange law similar to Concurrent Kleene Algebra. 
Each of these pairs satisfies a common biquantale structure. 
Additional structure is added via compatible sets of commands, including tests, atomic commands and pseudo-atomic commands.
These allow stronger (equality) interchange and distributive laws.
This paper describes the result of restructuring the algebra to better exploit these commonalities. 
The algebra is implemented in Isabelle/HOL.
\end{abstract}

\section{Introduction}

\paragraph{Motivation.}

Our end goal is to provide a foundation for deriving shared-variable, concurrent programs from specifications
expressed using the rely/guarantee approach of Jones \cite{Jones81d,Jones83a,Jones83b}.
His approach extends Floyd/Hoare style pre and post conditions \cite{Floyd67,Hoare69a} with rely and guarantee conditions.
When reasoning about a thread $T$, 
one needs to allow for the interference imposed on $T$ by its \emph{environment}, 
that is, the set of threads running in parallel with $T$.
A \emph{rely condition}, $r$, a binary relation between program states, 
represents an assumption that all transitions from a state $\sigma$ to a state $\sigma'$ made by the environment of $T$
satisfy $r$, that is, $(\sigma,\sigma') \in r$.
Complementing this, a thread ensures that all transitions it makes satisfy a \emph{guarantee condition}, $g$,
also a binary relation between states.
The guarantee condition of a thread $T$ must imply the rely conditions of all threads that run in parallel with $T$.

\paragraph{Semantic model.}

\begin{figure}
\begin{center}
\input{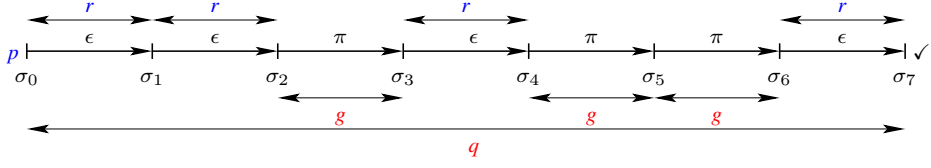}
\end{center}

\caption{Aczel trace satisfying a specification with precondition $p$, postcondition $q$, rely condition $r$ and guarantee condition $g$.
If the initial state, $\sigma_0$, is in $p$ and all environment transitions ($\estepd$) satisfy $r$,
then all program transitions ($\pstepd$) must satisfy $g$ and 
the initial state $\sigma_0$ must be related to the final state $\sigma_7$ by the postcondition $q$, also a relation between states.
}\labelfig{rely-guar}
\end{figure}

To handle interference, the semantics of a command is represented by a set of Aczel traces \cite{Aczel83,DaSMfaWSLwC},
which explicitly record not only state-to-state transitions made by a thread $T$, called \emph{program} ($\pstepd$) transitions,
but also transitions made by the environment of $T$, called \emph{environment} ($\estepd$) transitions -- see \reffig{rely-guar}.
Program and environment transitions are \emph{atomic}, that is, they are indivisible.
Preconditions and rely conditions both represent assumptions: 
if a precondition fails to hold initially any behaviour is allowed,
and if an environment transition fails to satisfy a rely condition, any behaviour is allowed from that point on.
Failure of a precondition or a rely condition corresponds to Dijkstra's abort command \cite{Dijkstra75,Dijkstra76}. 
Traces in the semantic model may be either terminating, aborting or incomplete.
A set of traces representing a command is prefix closed, 
that is, for any trace $tr$ in the set, all incomplete prefixes of $tr$ are in the set,
and sets of traces are abort closed,
that is, for any aborting trace in the set, all possible extensions of the aborting trace are in the set.
This semantic model supports rely/guarantee concurrency and is a model of our concurrent refinement algebra.

\refsect{structure} introduces the algebraic structures of monoids and quantales, along with iteration and fixed points.
\refsect{biquantale} combines pairs of quantales to form biquantales.
\refsect{closed} introduces compatible sets of commands,
of which 
tests (\refsect{tests}), 
atomic commands (\refsect{atomic}) and 
pseudo-atomic commands (\refsect{pseudo_atomic}) 
are instances.
\refsect{rely_guar} applies the above theories to rely/guarantee concurrency.

\section{Algebraic structure}\labelsect{structure}

The concurrent refinement algebra (CRA) \cite{AFfGRGRACP,FM2016atomicSteps,FMJournalAtomicSteps} 
is based on a complete lattice of commands, $\Command$,
with the lattice partial order, $c \refsto d$, corresponding to command $c$ is refined (or implemented) by command $d$, so that 
the lattice supremum ($\Nondet$) corresponds to non-deterministic choice,
the lattice infimum ($\Meet$) corresponds to a strong form of (specification) conjunction,
the least element of the lattice ($\bot$) corresponds to the everywhere infeasible command 
(called magic or miracle in the refinement calculus literature \cite{BackWright98,Morgan94}),
and
the greatest element ($\top$) corresponds to Dijkstra's abort command \cite{Dijkstra75,Dijkstra76}, 
that irrecoverably fails and hence allows any behaviour whatsoever.
Binary choice ($\nondet$) and strong conjunction ($\meet$) are defined in terms of $\Nondet$ and $\Meet$, respectively.
\begin{eqncolumns}
  c_1 \nondet c_2 & \defs \Nondet \{c_1,c_2\}  \labeldef{nondet}
\secondcolumn
  c_1 \meet c_2 & \defs \Meet \{c_1,c_2\} \labeldef{meet}
\end{eqncolumns}

\subsection{Monoid structure}\labelsect{monoid}

There are three primitive binary operations on the lattice of commands:
sequential composition, $c \Seq d$,
parallel composition, $c \parallel d$,
and
weak conjunction, $c \together d$,
that behaves as both $c$ and $d$ unless one of them aborts, in which case $c \together d$ aborts.%
\footnote{We discuss weak conjunction in more detail in \refsect{rely_guar}, for the moment one only needs to understand its algebraic properties.}

\begin{definition}[monoid]
A \emph{monoid}, $(S, \odot,\odotneut)$, consists of a set of elements, $S$, 
with an associative binary operation $\odot$ on $S$ with a neutral element $\odotneut \in S$,
such that $\odotneut \odot c = c = c \odot \odotneut$.
\end{definition}
The binary operations of our theory forms moniods, some of which are commutative and some also idempotent.
\begin{itemize}
\item
Sequential composition, $c \Seq d$, forms a monoid with neutral element the no-operation command, $\Nil$,
that is $\Nil \Seq c = c = c \Seq \Nil$.
\item
Parallel composition, $c \parallel d$, forms a commutative monoid with neutral element command $\Skip$,
that is, $\Skip \parallel c = c = c \parallel \Skip$.
\item
Weak conjunction, $c \together d$, forms a commutative, idempotent monoid%
\footnote{Equivalently, a semi-lattice with a neutral element, which is the structure available in Isabelle/HOL.}
with neutral element $\Chaos$,
that is, $\Chaos \together c = c = c \together \Chaos$.
\end{itemize}

\subsection{Quantale structure}

A (standard) quantale, $(Q,\odot)$, consists of a complete lattice, $Q$, with an associative binary operation, $\odot$,  on $Q$
satisfying the distributive laws \cite{2018StruthQuantales},
\begin{eqncolumns}
  d \odot (\Nondet C) & = \Nondet_{c \in C} (d \odot c) \labelax{quantale_left_distrib}
\secondcolumn
  (\Nondet C) \odot d & = \Nondet_{c \in C} (c \odot d)  \labelax{quantale_right_distrib}
\end{eqncolumns}
for $d \in Q$ and $C \subseteq Q$.
For the concurrent refinement algebra (CRA) these laws do not necessarily hold if $C$ is empty, 
because $\Nondet \{\} = \bot$.
In particular, for sequential composition \refax{quantale_left_distrib} requires, {\color{purple}$\top \Seq \bot = \bot$},
but because $\top$ is interpreted as irrecoverable failure (abort) in CRA, we require $\top \Seq \bot = \top$
because once a command has aborted not even a miracle can restore it.
Similar issues apply for $\parallel$ and $\together$ 
because in CRA both $\top \parallel c = \top$ and $\top \together c = \top$, for any command $c$, including $\Magic$.
For these reasons, we make use of a \emph{weak quantale} \cite{1989Paseka} 
that relaxes the above laws so they only need to hold for non-empty $C$.
 
\begin{definitionx}[quantales]
A \emph{pre-quantale} is a complete lattice with an associative binary operation $\odot$.

A \emph{weak left quantale} is a pre-quantale that distributes $\odot$ from the left over non-empty choices,
that is, for a command $d$ and non-empty set of commands $C$,
\begin{align}
  d \odot (\Nondet_{c \in C} c) & = \Nondet_{c \in C} (d \odot c) & \mbox{if } C \neq \{\} .  \labelax{op_distrib_Nondet} 
\end{align}
A \emph{weak right quantale} is a pre-quantale that distributes $\odot$ from the right over a non-empty set of commands,
\begin{align}
  (\Nondet_{c \in C} c) \odot d & = \Nondet_{c \in C} (c \odot d) & \mbox{if } C \neq \{\} .  \labelax{Nondet_distrib_op}
\end{align}
A \emph{weak quantale} is both a weak left quantale and a weak right quantale.

A \emph{left quantale} is a weak left quantale that distributes over empty choices, 
and hence because $\Nondet \emptyset = \Magic$, we have,
\(
  d \odot \bot = \bot .
\)

A \emph{right quantale} is a weak right quantale that distributes over empty choices, so that,
\(
 \bot \odot d = \bot .
\)

A \emph{quantale} is both a left and right quantale.

Any of the above quantales are termed \emph{unital} if they have a neutral element $\odotneut$,
as for a monoid.
\end{definitionx}

The distributive properties \refprop{op_distrib_Nondet} and \refprop{Nondet_distrib_op} 
are derived from \refax{op_distrib_Nondet} and \refax{Nondet_distrib_op}, respectively.
\begin{align}
  d \odot (c_1 \nondet c_2) & = (d \odot c_1) \nondet (d \odot c_2) \labelprop{op_distrib_Nondet} \\
  (c_1 \nondet c_2) \odot d & = (c_1 \odot d) \nondet (c_2 \odot d) \labelprop{Nondet_distrib_op} 
\end{align}
From \refprop{op_distrib_Nondet} and \refprop{Nondet_distrib_op}, 
$\odot$ is monotone with respect to refinement in its right and left operands, respectively.

One aspect of the restructuring of CRA is to make the common quantale structure of our theories explicit.
\begin{itemize}
\item
Sequential composition forms both a unital right quantale and a weak left quantale,
for which abort is a left annihilator for sequential composition \refax{abort_seq_annihilator}.
\item
Parallel composition forms a unital weak quantale, for which $\parallel$ is commutative,
and abort is an annihilator for parallel composition \refax{abort_par_annihilator}.
\item
Weak conjunction forms a unital weak quantale, for which $\together$ is commutative and idempotent,
and abort is an annihilator for weak conjunction \refax{abort_conj_annihilator}.
\end{itemize}
\begin{eqncolumns}
  \Abort \Seq c & = \Abort \labelax{abort_seq_annihilator} \\
  \Abort \parallel c & = \Abort \labelax{abort_par_annihilator} 
\secondcolumn
  \Abort \together c & = \Abort \labelax{abort_conj_annihilator}
\end{eqncolumns}

\subsection{Iteration and fixed points}\labelsect{iteration}

Fixed iteration, $c^i$ to a natural number power $i$ can be defined inductively on a quantale.%
\footnote{For our application in CRA, we only use iteration for sequential composition
but note that it could also be applied to parallel composition (e.g.\ $c^{\parallel 3} = c \parallel c \parallel c$)
but it is not useful for idempotent operations like $\together$ (because $c^{\together 3} = c \together c \together c = c$).}%
\begin{eqncolumns}
  c^0 & = \odotneut  \labelax{iter_zero} 
\secondcolumn
  c^{i+1} & = c \odot c^i \labelax{iter_succ}
\end{eqncolumns}
The least fixed point, $\mu f$, of a function $f$, is a fixed point \refax{mu_unfold} and is the least such \refax{mu_least} \cite{fixedpointcalculus1995}.
The greatest fixed point, $\nu f$, of a function $f$, is a fixed point \refax{nu_unfold} and is the greatest such \refax{nu_greatest}.
We abbreviate $\mu(\lambda y \spot f\,y)$ by $(\mu y \spot f\,y)$ and similarly, $\nu(\lambda y \spot f\,y)$ by $(\nu y \spot f\,y)$.
\begin{eqncolumns}
  \mu f & = & f (\mu f)  \labelax{mu_unfold} \\
  y & \geq & \mu f  ~~~~~~~~\mbox{if } y \geq f\, y \labelax{mu_least} 
\secondcolumn
  \nu f & = & f (\nu f)  \labelax{nu_unfold} \\
  \nu f & \geq & y ~~~~~~~~\mbox{if } f\,y \geq y \labelax{nu_greatest}
\end{eqncolumns}
Finite iteration, $\Fin{c}$, of a command $c$ is defined as the least fixed point  $(\mu y \spot \odotneut \nondet (c \odot y))$ \cite{Wright04}.
That gives both an unfolding property \refprop{finite_iter_unfold} and an induction property \refprop{finite_iter_induct}.%
\footnote{These correspond to axioms for Kleene star in Kleene algebra \cite{kozen97kleene}.}
Possibly infinite iteration, $\Om{c}$, is defined as the greatest fixed point $(\nu y \spot \odotneut \nondet (c \odot y))$,
giving an unfolding property \refprop{iter_unfold} and an induction property \refprop{iter_induct}.
Infinite iteration, $\Inf{c}$, is defined as the greatest fixed point $(\nu y \spot c \odot y)$.
That gives both an unfolding property \refprop{inf_iter_unfold} and an induction property \refprop{inf_iter_induct}.
\begin{eqncolumns}[40]
  \Fin{c} & = & \odotneut \nondet (c \odot \Fin{c}) \labelprop{finite_iter_unfold} \\
  \Om{c} & = & \odotneut \nondet (c \odot \Om{c}) \labelprop{iter_unfold} \\
  \Inf{c} & = & c \odot \Inf{c} \labelprop{inf_iter_unfold} 
\secondcolumn
  y & \refsto &\Fin{c} \odot d ~~~~\mbox{if } y \refsto d \nondet (c \odot y) \labelprop{finite_iter_induct} \\
  \Om{c} \odot d & \refsto & y ~~~~~~~~~~~~~\mbox{if } d \nondet (c \odot y) \refsto y \labelprop{iter_induct} \\
  \Inf{c} & \refsto & y ~~~~~~~~~~~~~\mbox{if } c \odot y \refsto y \labelprop{inf_iter_induct}
\end{eqncolumns}
\begin{lemmax}[finite-to-fixed] 
For all $i \in \nat$, $\Fin{c} \refsto c^i$.
\end{lemmax}

\begin{proof}
The proof is by induction on $i$.
For $i = 0$, using \refprop{finite_iter_unfold}, 
$\Fin{c} 
= \odotneut \nondet (c \odot \Fin{c}) 
\refsto \odotneut
= c^0
$.
Assuming the property for $i$, we can show it holds for $i+1$ as follows:
$\Fin{c} 
= \odotneut \nondet (c \odot \Fin{c}) 
\refsto c \odot \Fin{c}
\refsto c \odot c^i
= c^{i+1}
$.
\end{proof}
Because $\Fin{c} \refsto c^i$ for all $i \in \nat$, finite iteration is refined by a choice over all possible fixed iterations.
\begin{corx}[finite-to-choice]
$\Fin{c} \refsto \nondet_{i \in \nat} c^i$
\end{corx}
The reverse direction also holds if we have a weak left quantale, i.e.\ \refax{op_distrib_Nondet} holds.
\begin{lemmax}[finite-iter-decompose]
Assuming \refax{op_distrib_Nondet}, $\Fin{c} = \nondet_{i \in \nat} c^i$.
\end{lemmax}

\begin{proof}
From \refcor{finite-to-choice} it is sufficient to show refinement from right to left,
which holds by finite iteration induction \refprop{finite_iter_induct},
if $\Nondet_{i \in \nat} c^i \refsto \odotneut \nondet (c \odot \Nondet_{i \in \nat} c^i)$, 
which can be shown by pulling out $c^0$, which equals $\odotneut$ \refax{iter_zero}, 
and distributing using \refax{op_distrib_Nondet}, noting that $\nat$ is non-empty:
$\Nondet_{i \in \nat} c^i 
  = c^0 \nondet \Nondet_{i \in \nat} (c \odot c^i)
  = \odotneut \nondet (c \odot \Nondet_{i \in \nat} c^ i)
$.
\end{proof}

\section{Bi-quantale structure}\labelsect{biquantale}

\begin{figure}
\begin{center}
\input{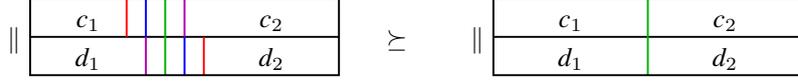}
\caption{Weak interchange law between parallel and sequential.
On the left $c_1$ may finish at any of the coloured lines and $d_1$ finish at the correspondingly coloured lines.
Hence $c_2$'s execution may overlap with $d_1$ or $c_1$'s execution may over lap with $d_2$.
On the right the termination of $c_1$ is synchronised with the termination of $d_1$.}\labelfig{interchange}
\end{center}
\end{figure}

Parallel and sequential composition satisfy a weak%
\footnote{It is termed \emph{weak} because it is only a refinement and not an equality.}
interchange law \refax{par_interchange_seq}, 
as in Concurrent Kleene Algebra \cite{DBLP:journals/jlp/HoareMSW11}.
\begin{align}
  (c_1 \Seq c_2) \parallel (d_1 \Seq d_2) & \refsto (c_1 \parallel d_1) \Seq (c_2 \parallel d_2) \labelax{par_interchange_seq} 
\end{align}
It reflects the fact that running $(c_1 \Seq c_2)$ in parallel with $(d_1 \Seq d_2)$ 
can be implemented by running $c_1$ in parallel with $d_1$ followed by running $c_2$ in parallel with $d_2$.
Note that the left side also allows, for example, 
$c_1$ to run in parallel with $d_1$ and part of $d_2$ and then $c_2$ to run in parallel with the rest of $d_2$,
or vice versa;
such behaviours are not permitted by the right side (see \reffig{interchange}).

As well as the weak interchage law for parallel and sequential \refax{par_interchange_seq}, 
the concurrent refinement algebra satisfies weak interchange laws for 
weak conjunction and sequential \refax{conj_interchange_seq},
and
for weak conjunction and parallel \refax{conj_interchange_par}.
The weak interchange law combining weak 
conjunction with sequential composition
\refax{conj_interchange_seq}
is similar to that for 
parallel and sequential composition \refax{par_interchange_seq}.
We defer explanation of \refax{conj_interchange_par} to \refsect{rely_guar}.
\begin{align}
  (c_1 \Seq c_2) \together (d_1 \Seq d_2) & \refsto (c_1 \together d_1) \Seq (c_2 \together d_2) \labelax{conj_interchange_seq} \\
  (c_1 \parallel c_2) \together (d_1 \parallel d_2) & \refsto (c_1 \together d_1) \parallel (c_2 \together d_2) \labelax{conj_interchange_par}
\end{align}

The weak interchange laws above may be abstracted to a single law \refax{otimes_interchange_odot} with abstract operations 
$\odot$ with neutral element $\odotneut$ and $\otimes$ with neutral element $\otimesneut$.
\begin{align}
  (c_1 \odot c_2) \otimes (d_1 \odot d_2) & \refsto (c_1 \otimes d_1) \odot (c_2 \otimes d_2) \labelax{otimes_interchange_odot}
\end{align}
Applying \refax{otimes_interchange_odot} and then using the properties of the neutral elements,
\begin{align*}&
  (\odotneut \odot \otimesneut) \otimes (\otimesneut \odot \odotneut) \refsto (\odotneut \otimes \otimesneut) \odot (\otimesneut \otimes \odotneut)
 \Equiv
  \otimesneut \otimes \otimesneut \refsto \odotneut \odot \odotneut
 \Equiv
  \otimesneut \refsto \odotneut . \numberthis\labelprop{iota-to-eta}
\end{align*}
Applying \refprop{iota-to-eta} to each instance, one has that 
$\Skip \refsto \Nil$ for ($\parallel$ and $\Seq$)
and 
$\Chaos \refsto \Nil$ for ($\together$ and $\Seq$)
and 
$\Chaos \refsto \Skip$
for ($\together$ and $\parallel$),
or combining these $\Chaos \refsto \Skip \refsto \Nil$, as expected in our intended model.
Using \refprop{iota-to-eta} one can deduce,
$
  \odotneut = \odotneut \otimes \otimesneut \refsto \odotneut \otimes \odotneut
$, that is,
\begin{align}
  \odotneut \refsto \odotneut \otimes \odotneut  . \labelprop{eta_to_eta_otimes_eta}
\end{align}
Applying \refprop{eta_to_eta_otimes_eta} to each instance, one has
$\Nil \refsto \Nil \parallel \Nil$
and
$\Nil \refsto \Nil \together \Nil$%
\footnote{\label{idem}This is already known to be an equality because $\together$ is idempotent.}
and 
$\Skip \refsto \Skip \together \Skip$,%
\footnote{This is already known to be an equality because $\together$ is idempotent.}
however, for our intended model these should all be equalities, so we introduce an additional axiom,
\begin{align}
  \odotneut \otimes \odotneut \refsto \odotneut \labelax{eta_otimes_eta_to_eta}
\end{align}
Similarly, using \refprop{iota-to-eta} one has,
$
  \otimesneut \odot \otimesneut \refsto \otimesneut \odot \odotneut = \otimesneut 
$, that is,
\begin{align}
  \otimesneut \odot \otimesneut \refsto \otimesneut \labelprop{iota_odot_iota_to_iota}
\end{align}
Applying \refprop{iota_odot_iota_to_iota} to each instance, one has
$\Skip \Seq \Skip \refsto \Skip$ for ($\parallel$ and $\Seq$)
and
$\Chaos \Seq \Chaos \refsto \Chaos$ for ($\together$ and $\Seq$)
and
$\Chaos \parallel \Chaos \refsto \Chaos$ for ($\together$ and $\parallel$),
however, for our intended model these should all be equalities, so we introduce an additional axiom,
\begin{align}
  \otimesneut \refsto \otimesneut \odot \otimesneut . \labelax{iota_to_iota_odot_iota}
\end{align}

We are now in a position to define \emph{biquantale} which captures the above properties.
In category theory, a biquantale is a 2-monoidal category \cite{2010AgularMahajanMonoidal}.
\begin{definitionx}[biquantale]
A biquantale $(Q,\otimes, \otimesneut,\odot,\odotneut)$, consists of a complete lattice $Q$
with a binary operation $\odot$ with neutral element $\odotneut$, 
such that $(Q,\odot,\odotneut)$ forms a unital weak quantale,
and a commutative binary operation $\otimes$ with neutral element $\otimesneut$, 
such that $(Q,\otimes,\otimesneut)$ forms a unital weak quantale,
and together they satisfy axioms \refax{otimes_interchange_odot}, \refax{eta_otimes_eta_to_eta} and \refax{iota_to_iota_odot_iota}.
\end{definitionx}
If $(Q,\odot,\odotneut)$ is a quantale,
then both $(Q,\meet,\top,\odot,\odotneut)$ and $(Q,\odot,\odotneut,\nondet,\bot)$ form biquantales.

As part of our restructuring we have made the common biquantale structure explicit for each of our pairs of operations.
\begin{definitionx}[General Concurrent Refinement Algebra]
A \emph{general concurrent refinement algebra} consists of 
\begin{itemize}
\item
a biquantale with $\parallel$ for $\sync$ and sequential composition ($\Seq$) for $\odot$,
in which sequential composition is a right quantale,
\item
a biquantale with $\together$ for $\sync$ and sequential composition ($\Seq$) for $\odot$,
in which $\together$ is idempotent and sequential composition is a right quantale, 
and
\item
a biquantale with weak conjunction for $\sync$ and parallel composition for $\odot$,
in which parallel composition is commutative and weak conjunction is idempotent.
\end{itemize}
If a command $c_1$ cannot synchronise with with $\Nil$, any extension of it cannot \refax{whole_if_first};
this axiom applies with $\sync$ both $\parallel$ and $\together$.
\begin{align}
  \Nil \sync (c_1 \Seq c_2) & = \Magic & \mbox{if } \Nil \sync c_1 = \Magic \labelax{whole_if_first}
\end{align}
\end{definitionx}

\subsection{Distributive laws in a biquantale}\labelsect{biquantale_distrib}

From the weak interchange law \refax{otimes_interchange_odot}, if $d \refsto d \odot d$, one can deduce
$d \sync (c_1 \odot c_2) \refsto (d \odot d) \sync (c_1 \odot c_2) \refsto (d \sync c_1) \odot (d \sync c_2)$,
that is,
\begin{align}
  d \sync (c_1 \odot c_2) & \refsto (d \sync c_1) \odot (d \sync c_2) .  \labelprop{weak_sync_distrib_odot}
\end{align}
As in Concurrent Kleene Algebra \cite{DBLP:journals/jlp/HoareMSW11},
as well as \refprop{weak_sync_distrib_odot} the following laws hold for iterations.
\begin{align}
  d \sync c^0 & \refsto (d \sync c)^0 & \mbox{if } d \sync \odotneut \refsto \odotneut \labelprop{weak_sync_distrib_iter_zero} \\
  d \sync c^{i+1} & \refsto (d \sync c)^{i+1} & \mbox{if } d \refsto d \odot d \labelprop{weak_sync_distrib_iter_succ} \\
  d \sync \Fin{c} & \refsto \Fin{(d \sync c)} & \mbox{if } d \sync \odotneut \refsto \odotneut \mbox{ and }d \refsto d \odot d \labelprop{weak_sync_distrib_finite_iter} 
\end{align}
If the following strengthening of \refprop{weak_sync_distrib_odot} holds,
\begin{align}
  d \sync (c_1 \odot c_2) & = (d \sync c_1) \odot (d \sync c_2) ,  \labelprop{sync_distrib_odot}
\end{align}
the distributive laws for iterations may be strengthened to the following equalities.
\begin{align}
  d \sync c^0 & = (d \sync c)^0 & \mbox{if } d \sync \odotneut = \odotneut \labelprop{sync_distrib_iter_zero} \\
  d \sync c^{i+1} & = (d \sync c)^{i+1} & \mbox{if \refprop{sync_distrib_odot} for all $c_1$ and $c_2$} \labelprop{sync_distrib_iter_succ} \\
  d \sync \Fin{c} & = \Fin{(d \sync c)} & \mbox{if } d \sync \odotneut = \odotneut \mbox{ and \refprop{sync_distrib_odot} for all $c_1$ and $c_2$} \labelprop{sync_distrib_finite_iter} 
\end{align}
The first \refprop{sync_distrib_iter_zero} follows directly from the assumption and the definition of iteration to the power zero \refax{iter_zero};
\refprop{sync_distrib_iter_succ} can be shown by induction on $i$ using \refprop{sync_distrib_odot}; 
and
\refprop{sync_distrib_finite_iter} follows from \reflem{finite-iter-decompose},
distributing using \refax{op_distrib_Nondet},
and applying \refprop{sync_distrib_iter_zero} and \refprop{sync_distrib_iter_succ}:
$d \sync \Fin{c} 
= d \sync \Nondet_{i \in \nat} c^i
= \Nondet_{i \in \nat} (d \sync c^i)
= \Nondet_{i \in \nat} (d \sync c)^i
= \Fin{(d \sync c)}
$.
\refsect{atomic} examines forms for the command $d$ that satisfy the assumptions for these properties, 
in particular, \refprop{sync_distrib_odot}.

\section{Compatible sets of commands}\labelsect{closed}

As part of our restructure we introduced the notion of a compatible set of elements.
In CRA, the set of test commands (see \refsect{tests}) and the set of atomic commands (see \refsect{atomic}) both form compatible sets.
\begin{definitionx}[compatible]
A subset, $\Compatible$, of a biquantale $Q$ is \emph{compatible} if it is closed under 
arbitrary suprema \refax{closed_sup_closed}, 
non-empty infima \refax{closed_inf_closed}
and under the binary operator $\otimes$ \refax{closed_sync_closed}.
In addition, an arbitrary command $c$ distributes from the right into an infimum of a set of compatible commands \refax{closed_odot_distrib_Meet_right}.
\begin{align}
  Z \subseteq \Compatible & \implies \Nondet Z \in \Compatible \labelax{closed_sup_closed} \\
  Z \subseteq \Compatible \land Z \neq \{\} & \implies \Meet Z \in \Compatible \labelax{closed_inf_closed} \\
  c_1 \in \Compatible \land c_2 \in \Compatible & \implies c_1 \sync c_2 \in \Compatible \labelax{closed_sync_closed} \\
  Z \subseteq \Compatible & \implies (\Meet Z) \odot c = \Meet_{z \in Z} (z \odot c) \labelax{closed_odot_distrib_Meet_right}
\end{align}
\end{definitionx}
Because the set of commands is based on a quantale, we already have that non-empty suprema distribute over $\odot$ \refax{Nondet_distrib_op}.
We emphasise that \refax{closed_odot_distrib_Meet_right} does not hold for an arbitrary set of commands $Z$;
in CRA it requires all the commands in $Z$ make the same number of transitions,
e.g.\ $Z$ is a set of tests, which all take no transitions (see \refsect{tests})
or $Z$ is a set of atomic commands, which all take a single transitions (see \refsect{atomic}).

\section{Tests}\labelsect{tests}

In order to define conditionals, we introduce a set of instantaneous tests, $\Test$, 
similar to the tests in Kozen's Kleene Algebra with tests \cite{kozen97kleene}.
The set of instantaneous test commands, $\Test$, forms a compatible set of commands, that is,
$\Test$ is closed under non-deterministic choice, non-empty infima and $\otimes$,
and a command distributes via $\odot$ from the right over a conjunction of tests.
All tests refine the neutral element $\odotneut$ of $\odot$ \refax{eta_to_test},
which implies that $\odotneut$ is the greatest (true) test and $\bot$ is the least (false) test.
The negation, $\lnot t$ of a test is a test \refax{test_negate}, 
and tests satisfy the additional axioms \refax{test_and_negate_disjoint} and \refax{test_or_negate_univ} 
that ensure $\Test$ forms a boolean algebra.
In addition, the operations $\odot$ and $\otimes$ applied to a pair of tests 
correspond to their strong conjunction, \refax{test_odot_test} and \refax{test_sync_test},
so tests are also closed under $\odot$ (because they are closed under strong conjunction). 
A test distributes into a synchronisation \refax{test_distrib_sync}.
For $t, t_1, t_2 \in \Test$,
\begin{eqncolumns}[40]
  t \in \Test & \iff & \odotneut \refsto t  \labelax{eta_to_test} \\
  t \in \Test & \iff & \lnot t \in \Test \labelax{test_negate} \\
  t \meet \lnot t & = & \bot \labelax{test_and_negate_disjoint} \\
  t \nondet \lnot t & = & \odotneut \labelax{test_or_negate_univ}
\secondcolumn
  t_1 \odot t_2 & = & t_1 \meet t_2 \labelax{test_odot_test} \\
  t_1 \sync t_2 & = & t_1 \meet t_2 \labelax{test_sync_test} \\
  t \odot (c_1 \sync c_2) & = & (t \odot c_1) \sync (t \odot c_2)  \labelax{test_distrib_sync} 
\end{eqncolumns}\\
A final test on either branch of a synchronisation may be pulled out as a final test of the whole synchronisation \refprop{final_test} \cite{hayes2021deriving}.
\begin{align}
  (c_1 \odot t) \sync c_2 & = (c_1 \sync c_2) \odot t  \labelprop{final_test}
\end{align}

\subsection{Assertions}\labelsect{assertions}

In order to encode preconditions as a command, we introduce an assert command, $\Assert{t}$,
that is defined in terms of a test $t$.
While a test $t$ is infeasible if the test $t$ does not hold for the initial state,
the assertion command, $\Assert{t}$, aborts if $t$ does not hold \refdef{assert}.
An equivalent definition is \refprop{assert-alt}.
An assertion followed by a command behaves as the command unless $t$ does not hold, in which case it aborts \refprop{assert-seq}.
\begin{align}
  \Assert{t} & \defs \odotneut \nondet (\lnot t \odot \Abort) \labeldef{assert} \\
  \Assert{t} & = t \nondet (\lnot t \odot \Abort) \labelprop{assert-alt} \\
  \Assert{t} \odot c & = c \nondet (\lnot t \odot \Abort) \labelprop{assert-seq}
\end{align}

\section{Atomic algebra}\labelsect{atomic}

In Aczel traces \cite{Aczel83,DaSMfaWSLwC}, program and environment transitions are indivisible or \emph{atomic} (see \reffig{rely-guar}).
We can reflect this abstractly in the algebra by introducing a compatible subset of atomic commands, $\Atomic$.
For rely/guarantee concurrency these will be instantiated as commands that can make 
a single program or environment transition and then terminate but other instantiations are possible (see \refsect{pseudo_atomic}).
Notably $\Atomic$ is not closed under sequential composition
(because the sequential composition of two atomic commands may make two transitions).
If the initial commands in an interchange are atomic, then the interchange law is strong, 
that is, an equality rather than a refinement \refax{sync_interchange_odot}.
An atomic command cannot synchronise with $\odotneut$ because $\odotneut$ cannot make any transitions \refax{atomic_sync_eta}.
The atomic command $\atomicotimesneut$ is the neutral element for $\sync$ for atomic commands \refax{atomic_neutral}.
Using induction on $i$ one can show $\Ata_1^i \sync \Ata_2^i = (\Ata_1 \sync \Ata_2)^i$ using \refax{sync_interchange_odot}.
Axiom \refax{atomic_interchange_inf_iter} extends that to infinite iterations.

\begin{definition}[atomic algebra]
A subset, $\Atomic$, of elements over a biquantale forms an \emph{atomic algebra} 
if $\Atomic$ forms a compatible set of commands and the following axioms hold.
For $\Ata, \Ata_1, \Ata_2 \in \Atomic$,
\begin{eqncolumns}[70]
  (\Ata_1 \odot c_1) \sync (\Ata_2 \odot c_2) &= & (\Ata_1 \sync \Ata_2) \odot (c_1 \sync c_2) \labelax{sync_interchange_odot} \\
  \Inf{\Ata_1} \sync \Inf{\Ata_2} & = & \Inf{(\Ata_1 \sync \Ata_2)} \labelax{atomic_interchange_inf_iter}
\secondcolumn
  \odotneut \sync \Ata & = & \bot \labelax{atomic_sync_eta} \\
  \atomicotimesneut \sync \Ata & = & \Ata \labelax{atomic_neutral}
\end{eqncolumns}\\
In addition, any command, $c$, can either 
abort if some test $t'$ holds,
terminate immediately if some test $t$ holds,
or make a transition of an atomic command $\Ata$ and 
then behave as $c'$ for some set of pairs $C \subseteq \Atomic \times \Command$ of atomic command and command, that is,
\begin{align}
  \exists t', t, C \spot c = \Assert{t'} \Seq (t \nondet \Nondet_{(\Ata,c') \in C} (\Ata \Seq c')) . \labelax{expanded_form}
\end{align}
\end{definition}
From \refax{atomic_sync_eta} and \refax{whole_if_first} we can derive the following property for $\Ata \in \Atomic$.
\begin{align}
  \odotneut \sync (\Ata \odot c) = \bot \labelprop{atomic_odot_sync_eta}
\end{align}
Because $\atomicotimesneut$ is the atomic neutral element for $\sync$, 
for an atomic algebra $\Om{\atomicotimesneut}$ is the general neutral element for $\sync$, 
that is $\otimesneut = \Om{\atomicotimesneut}$.

\subsection{Distributive laws with atomic commands}\labelsect{atomic_distrib}

A command $d$ is an \emph{atomic fixed point} if for some atomic command $\Ata$ it is a fixed point of the following equation,
\begin{align}
  d & = \odotneut \nondet (\Ata \odot d) . \labeldef{atomic_fixed_point}
\end{align}
From \refprop{finite_iter_unfold} and \refprop{iter_unfold}, 
if $\Ata$ is an atomic command then both $\Fin{\Ata}$ and $\Om{\Ata}$ are both examples of atomic fixed points.
Commands that are atomic fixed points satisfy the assumptions,
\begin{align}
  d \sync \odotneut & = \odotneut \labelprop{atomic_fp_sync_eta} \\
  d \sync (c_1 \odot c_2) & = (d \sync c_1) \odot (d \sync c_2) & \mbox{for all commands $c_1$ and $c_2$} \labelprop{atomic_fp_distrib_seq}
\end{align}
for the distribution of $d$ into 
fixed iterations (\refprop*{sync_distrib_iter_zero}--\refprop*{sync_distrib_iter_succ}),
and finite iterations \refprop{sync_distrib_finite_iter}.
The proof of \refprop{atomic_fp_sync_eta} follows by 
expanding $d$ using \refdef{atomic_fixed_point},
distributing,
and simplifying using \refprop{eta_to_eta_otimes_eta}, \refax{eta_otimes_eta_to_eta} and \refprop{atomic_odot_sync_eta}.
The proof of \refprop{atomic_fp_distrib_seq} is somewhat more complex:
it uses the fact that $c_1$ can be decomposed into its behaviours of length $i$ for $i \in \nat$, $c_1 \sync \atomicotimesneut^i$,
and its infinite behaviours, $c_1 \sync \Inf{\atomicotimesneut}$,
using the fact that $\atomicotimesneut$ is the atomic identity of $\sync$.
The distribution property is shown to hold for each of the components of $c_1$
before recomposing the resulting components.
The proofs of the finite cases make use of the fact that a command can be rewritten in expanded form using \refax{expanded_form}.
The reader is referred to \cite{2023MeinickeHayesDistributive-TR} for full details.

\subsection{Iteration with atomic commands}\labelsect{atomic_iteration}

The following properties hold if $\Ata_1$ and $\Ata_2$ are atomic commands.
Significantly, all these properties are equalities.
Their detailed proofs may be found in  \cite{FM2016atomicSteps,FMJournalAtomicSteps}.
\begin{align}
  (\Ata_1^i \odot c_1) \sync (\Ata_2^i \odot c_2) & = (\Ata_1 \sync \Ata_2)^i \odot (c_1 \sync c_2) \labelprop{atomic_sync_fixed_iter_prefix} \\
  \Ata_1^i \sync \Ata_2^i & = (\Ata_1 \sync \Ata_2)^i \labelprop{atomic_sync_fixed_iter} \\
  (\Fin{\Ata_1} \odot c_1) \sync (\Fin{\Ata_2} \odot c_2) & = \Fin{(\Ata_1 \sync \Ata_2)} \odot ((c_1 \sync (\Fin{\Ata_2} \odot c_2)) \nondet ((\Fin{\Ata_1} \odot c_1) \sync c_2)) \labelprop{atomic_sync_finite_iter_prefix} \\
  \Fin{\Ata_1} \sync \Fin {\Ata_2} & = \Fin{(\Ata_1 \sync \Ata_2)} \labelprop{atomic_sync_finite_iter} \\
  (\Fin{\Ata_1} \odot c_1) \sync \Inf{\Ata_2} & = \Fin{(\Ata_1 \sync \Ata_2)} \odot (c_1 \sync \Inf{\Ata_2}) \labelprop{atomic_sync_finite_iter_infinite} \\
  (\Om{\Ata_1} \odot c_1) \sync \Inf{\Ata_2} & = \Om{(\Ata_1 \sync \Ata_2)} \odot (c_1 \sync \Inf{\Ata_2}) \labelprop{atomic_sync_iter_infinite} \\
  (\Om{\Ata_1} \odot c_1) \sync (\Om{\Ata_2} \odot c_2) & = \Om{(\Ata_1 \sync \Ata_2)} \odot ((c_1 \sync (\Om{\Ata_2} \odot c_2)) \nondet ((\Om{\Ata_1} \odot c_1) \sync c_2)) \labelprop{atomic_sync_iter_prefix}  \\
  \Om{\Ata_1} \sync \Om{\Ata_2} & = \Om{(\Ata_1 \sync \Ata_2)} \labelprop{atomic_sync_iter}
\end{align}
Law \refprop{atomic_sync_fixed_iter_prefix} can be shown by induction on $i$ using \refax{sync_interchange_odot}, and 
\refprop{atomic_sync_fixed_iter} can be shown as a corollary by taking both $c_1$ and $c_2$ to be $\odotneut$.
Law \refprop{atomic_sync_finite_iter_prefix} is shown by 
decomposing the finite iterations into fixed iterations, $\Ata_1^i$ for $i \in \nat$ and $\Ata_2^j$ for $j \in \nat$ using \reflem{finite-iter-decompose},
and then considering the combinations when $i \leq j$ and $i \geq j$, separately.
Informally, $\Fin{\Ata_1}$ and $\Fin{\Ata_2}$ can synchronise some finite number of steps to give $\Fin{(\Ata_1 \sync \Ata_2)}$, then 
if $\Fin{\Ata_1}$ is exhausted, $c_1$ synchronises with the remaining steps of $\Fin{\Ata_2}$ followed by $c_2$,
or
if $\Fin{\Ata_2}$ is exhausted, $c_2$ synchronises with the remaining steps of $\Fin{\Ata_1}$ followed by $c_1$.
Law \refprop{atomic_sync_finite_iter} can be shown as a corollary by taking both $c_1$ and $c_2$ to be $\odotneut$.
Law \refprop{atomic_sync_finite_iter_infinite} has a proof similar to \refprop{atomic_sync_finite_iter_prefix}
but in this case $\Inf{\Ata_2}$ cannot terminate.
Law \refprop{atomic_sync_iter_infinite} is shown by splitting $\Om{\Ata_1}$ into $\Fin{\Ata_1}$ and $\Inf{\Ata_1}$,
and using \refprop{atomic_sync_finite_iter_infinite} to combine the finite component,
and \refax{atomic_interchange_inf_iter} to combine the infinite components.
Law \refprop{atomic_sync_iter_prefix} can be shown by splitting $\Om{\Ata_1}$ into $\Fin{\Ata_1}$ and $\Inf{\Ata_1}$,
and $\Om{\Ata_2}$ into $\Fin{\Ata_2}$ and $\Inf{\Ata_2}$,
using \refprop{atomic_sync_finite_iter_prefix} to combine the finite components,
\refprop{atomic_sync_finite_iter_infinite} to combine finite components with infinite components,
and \refax{atomic_interchange_inf_iter} to combine the infinite components.
Law \refprop{atomic_sync_iter} can be shown as a corollary of \refprop{atomic_sync_iter_prefix} by taking both $c_1$ and $c_2$ to be $\odotneut$.

\section{Pseudo-atomic commands}\labelsect{pseudo_atomic}

Given a set of atomic commands, $\Atomic$, one can define a set of pseudo-atomic commands, $\PseudoAtomic$,
each of the form, $\Ata \nondet (\Atb \odot \Abort)$, for some $\Ata$ and $\Atb$ in $\Atomic$.
A pseudo-atomic command can either perform atomic command $\Ata$ or it can perform atomic command $\Atb$ and then abort.
Interestingly, if $\Atomic$ forms an atomic algebra, then so does $\PseudoAtomic$,
that is, $\PseudoAtomic$ is a compatible set of commands and satisfies axioms (\refax*{sync_interchange_odot}--\refax*{atomic_interchange_inf_iter})
but with pseudo-atomic commands in place of atomic commands.
That means the distributive laws from \refsect{atomic_distrib} are also applicable to pseudo-atomic commands.
An atomic fixed point command becomes a pseudo-atomic fixed point that, 
for some pseudo-atomic command $\Patx$, satisfies,
\begin{align}
  d & = \odotneut \nondet (\Patx \odot d) . \labeldef{pseudo-atomic-fixed-point}
\end{align}
Hence pseudo-atomic fixed points, like $\Fin{\Patx}$ and $\Om{\Patx}$, 
satisfy \refprop{atomic_fp_sync_eta} and \refprop{atomic_fp_distrib_seq}
and hence distribute into fixed and finite iterations (\refprop*{sync_distrib_iter_zero}--\refprop*{sync_distrib_finite_iter}).
Further they satisfy the properties of iterations in \refsect{atomic_iteration} but with pseudo-atomic commands in place of atomic commands.

The realisation that pseudo-atomic commands form an atomic algebra saves repeating a significant number of proofs,
which have exactly the same structure as the proofs for atomic commands,
but are much more complicated 
because each atomic command is replaced by a more complex pseudo-atomic command of the form $\Ata \nondet (\Atb \odot \Abort)$.

\section{Application to rely/guarantee concurrency}\labelsect{rely_guar}

For rely/guarantee concurrency we instantiate the set of atomic commands, $\Atomic$,
with commands of the form, $\cpstep{g} \nondet \cestep{r}$, where $r$ and $g$ are binary relations between states and,
\begin{description}
\item[$\cpstep{g}$] is an \emph{atomic program} command 
that can perform a program ($\pstepd$) transition from state $\sigma$ to $\sigma'$
if $(\sigma, \sigma') \in g$, and
\item[$\cestep{r}$] is an \emph{atomic environment} command 
that can perform an environment ($\estepd$) transition from state $\sigma$ to $\sigma'$
if $(\sigma, \sigma') \in r$.
\end{description}
For these commands reducing the relation is a refinement,
that is, both $\cpstep{r_1} \refsto \cpstep{r_2}$ and $\cestep{r_1} \refsto \cestep{r_2}$ hold if $r_1 \supseteq r_2$.
Given that $\universalrel$ is the universal relation between program states, 
the command $\cpstepd$ (note the bold font) can perform any program transition \refdef{cpstepd},
$\cestepd$ any environment transition \refdef{cestepd},
and $\cstepd$ any transition, program or environment \refdef{cstepd}.
\begin{eqncolumns}
  \cpstepd & \defs \cpstep{\universalrel} \labeldef{cpstepd} \\
  \cestepd & \defs \cestep{\universalrel} \labeldef{cestepd} 
\secondcolumn
  \cstepd & \defs & \cpstepd \nondet \cestepd \labeldef{cstepd} 
\end{eqncolumns}
Weak conjunction of atomic commands is the same as their strong conjunction (the lattice meet) 
because they do not abort (\refax*{cpstep_conj_cpstep}--\refax*{cestep_conj_cestep}).
Program and environment commands are disjoint and hence their weak conjunction is infeasible \refax{cpstep_conj_cestep}.
Parallel composition of atomic commands matches a program transition of one command with an environment transition of the other
to give a program transition of the parallel composition \refax{cpstep_par_cestep}.
It matches environment transitions of both to give an environment transition of the parallel composition \refax{cestep_par_cestep}.
Two threads cannot both do a program transition at the same time, hence combining to atomic program commands is infeasible \refax{cpstep_par_cpstep};
this has the effect of interleaving program transitions of different threads.
\begin{eqncolumns}
  \cpstep{r_1 } \together \cpstep{r_2} & = & \cpstep{(r_1 \inter r_2)} \labelax{cpstep_conj_cpstep} \\
  \cestep{r_1 } \together \cestep{r_2} & = & \cestep{(r_1 \inter r_2)} \labelax{cestep_conj_cestep} \\
  \cpstep{r_1 } \together \cestep{r_2} & = & \Magic \labelax{cpstep_conj_cestep}
\secondcolumn
  \cpstep{r_1 } \parallel \cestep{r_2} & = & \cpstep{(r_1 \inter r_2)} \labelax{cpstep_par_cestep} \\
  \cestep{r_1 } \parallel \cestep{r_2} & = & \cestep{(r_1 \inter r_2)} \labelax{cestep_par_cestep} \\
  \cpstep{r_1 } \parallel \cpstep{r_2} & = &\Magic \labelax{cpstep_par_cpstep}
\end{eqncolumns}
The atomic neutral element of weak conjunction is the command, $\cstepd$, that allows any program or environment transition \refdef{cstepd}.
The neutral element of weak conjunction allows any number of such transitions \refdef{chaos}.
The atomic neutral element of parallel composition is the command, $\cestepd$, that allows any environment transition \refdef{cestepd}.
The neutral element of parallel composition allows any number of such transitions \refdef{skip}.
\begin{eqncolumns}
  \Chaos & \defs & \Om{\cstepd} \labeldef{chaos}
\secondcolumn
  \Skip & \defs & \Om{\cestepd} \labeldef{skip}
\end{eqncolumns}

\subsection{Guarantees}\labelsect{guar}

The guarantee command, $\guar{g}$, restricts program transitions to satisfy $g$ but puts no constraints on environment transitions.
\begin{align}
  \guar{g} & \defs \Om{(\cpstep{g} \nondet \cestepd)} \labeldef{guar}
\end{align}
As a guarantee is defined in terms of atomic commands we can use \refprop{atomic_sync_iter} to show,
\begin{align}
  \guar{g_1} \together \guar{g_2} & = \guar{(g_1 \inter g_2)} \labelprop{guar_conj} \\
  \guar{g_1} \parallel \guar{g_2} & = \guar{(g_1 \union g_2)} \labelprop{guar_par}
\end{align}
because distributing and using (\refax*{cpstep_conj_cpstep}--\refax*{cpstep_conj_cestep}) 
and (\refax*{cpstep_par_cestep}--\refax*{cpstep_par_cpstep}),
\begin{align*}
(\cpstep{g_1} \nondet \cestepd) \together (\cpstep{g_2} \nondet \cestepd) & = \cpstep{(g_1 \inter g_2)} \nondet \cestepd \\
(\cpstep{g_1} \nondet \cestepd) \parallel (\cpstep{g_2} \nondet \cestepd) & = \cpstep{(g_1 \union g_2)} \nondet \cestepd .
\end{align*}
Because a guarantee is an iteration of an atomic command, and hence it is an atomic fixed point,
a guarantee distributes over a sequential composition by \refprop{atomic_fp_distrib_seq}.
\begin{align}
  \guar{g} \together (c_1 \Seq c_2) & = (\guar{g} \together c_1) \Seq (\guar{g} \together c_2) \labelprop{guar_distrib-seq}
\end{align}

\subsection{Termination}\labelsect{term}

The command $\Term$ is the most general command that only performs a finite number of program transitions
but does not constrain its environment.
\begin{align}
  \Term & \defs \Fin{\cstepd} \Seq \Om{\cestepd} \labeldef{term}
\end{align}
Two terminating commands in parallel, terminate.
\begin{lemmax}[term-par-term]
$\Term \parallel \Term = \Term$
\end{lemmax}

\begin{proof}
The proof unfolds the definition of $\Term$, uses \refprop{atomic_sync_finite_iter_prefix},
the fact the $\Om{\cestepd}$ is the identity of parallel \refdef{skip},
and the fact that $\Fin{c} \Seq \Fin{c} = \Fin{c}$ for any command $c$,
before folding the definition of $\Term$:
$
  \Term \parallel \Term
= \Fin{\cstepd} \Seq \Om{\cestepd} \parallel \Fin{\cstepd} \Seq \Om{\cestepd}
= \Fin{(\cstepd \parallel \cstepd)} \Seq (\Om{\cestepd} \parallel \Fin{\cstepd} \Seq \Om{\cestepd})
= \Fin{\cstepd} \Seq \Fin{\cstepd} \Seq \Om{\cestepd}
= \Fin{\cstepd} \Seq \Om{\cestepd}
= \Term
$,
where it is straightforward to show $\cstepd \parallel \cstepd = \cstepd$.
\end{proof}

\subsection{Relies}\labelsect{relies}

The rely command, $\rely{r}$, puts no constraints on either program or environment transitions
but if the environment makes a transition not satisfying $r$, the rely command aborts.
\begin{align}
  \rely{r} & \defs \Om{(\cpstepd \nondet \cestepd \nondet \cestep{\overline{r}} \Seq \Abort)} \labeldef{rely}
\end{align}
As $r_1 \subseteq r_2$ implies $\overline{r_1} \supseteq \overline{r_2}$,
weakening the relation in a rely command is a refinement.
\begin{align}
  \rely{r_1} & \refsto \rely{r_2} & \mbox{if } r_1 \subseteq r_2 \labelprop{rely_weaken}
\end{align}

Because a rely is defined in terms of a pseudo-atomic commands we can use \refprop{atomic_sync_iter}, 
but on pseudo-atomic commands rather than atomic commands, to show,
\begin{align}
  \rely{r_1} \together \rely{r_2} & = \rely{(r_1 \inter r_2)} \labelprop{rely_conj} 
\end{align}
because distributing and using (\refax*{cpstep_conj_cpstep}--\refax*{cpstep_conj_cestep}), 
\refax{sync_interchange_odot} and \refax{abort_seq_annihilator},
\begin{align*}
& (\cpstepd \nondet \cestepd \nondet \cestep{\overline{r_1}} \Seq \Abort) \together (\cpstepd \nondet \cestepd \nondet \cestep{\overline{r_2}} \Seq \Abort) 
\\
& = \begin{array}[t]{l}
 	(\cpstepd \together \cpstepd) \nondet (\cpstepd \together \cestepd) \nondet (\cpstepd \together \cestep{\overline{r_2}} \Seq \Abort) \nondet {} \\
	(\cestepd \together \cpstepd) \nondet (\cestepd \together \cestepd) \nondet (\cestepd \together \cestep{\overline{r_2}} \Seq \Abort) \nondet {} \\
	(\cestep{\overline{r_1}} \Seq \Abort \together \cpstepd) \nondet (\cestep{\overline{r_1}} \Seq \Abort \together \cestepd) \nondet (\cestep{\overline{r_1}} \Seq \Abort \together \cestep{\overline{r_2}} \Seq \Abort)
      \end{array}
\\
 & =  (\cpstepd \nondet \Magic \nondet \Magic \Seq \Abort) \nondet 
	(\Magic \nondet \cestepd \nondet \cestep{\overline{r_2}} \Seq \Abort) \nondet 
	(\Magic \nondet \cestep{\overline{r_1}} \Seq \Abort \nondet \cestep{(\overline{r_1} \inter \overline{r_2})} \Seq \Abort)
\\
 & = \cpstepd \nondet \cestepd \nondet (\cestep{\overline{r_1}} \nondet \cestep{\overline{r_2}} \nondet \cestep{(\overline{r_1} \inter \overline{r_2})}) \Seq \Abort \\
 & =\cpstepd \nondet \cestepd \nondet \cestep{(\overline{r_1} \union \overline{r_2} \union (\overline{r_1} \inter \overline{r_2}))} \Seq \Abort \\
 & = \cpstepd \nondet \cestepd \nondet \cestep{(\overline{r_1 \inter r_2})} \Seq \Abort .
\end{align*}

A more interesting and not immediately obvious property is the following,
which we apply for developing the parallel introduction law below.
\begin{align}
  \rely{r} \parallel \guar{r} & = \rely{r} \labelprop{rely_par_guar}
\end{align}
Applying  \refprop{atomic_sync_iter} on pseudo-atomic commands,%
\footnote{Note that an atomic command, $\Ata$, is also a pseudo-atomic command 
because $\Ata = \Ata \nondet \Magic \Seq \Abort$ and $\Magic$ is an atomic command.}
it is sufficient to show,
\begin{align}
  (\cpstepd \nondet \cestepd \nondet \cestep{\overline{r}} \Seq \Abort) \parallel (\cpstep{r} \nondet \cestepd) & = \cpstepd \nondet \cestepd \nondet \cestep{\overline{r}} \Seq \Abort
\end{align}
which holds by distributing and applying (\refax*{cpstep_par_cestep}--\refax*{cpstep_par_cpstep}), 
noting that $\cpstep{(\overline{r} \inter r)} \Seq \Abort = \cpstep{\emptyset} \Seq \Abort =\Magic \Seq \Abort = \Magic$,
\begin{align*}
   & (\cpstepd \nondet \cestepd \nondet \cestep{\overline{r}} \Seq \Abort) \parallel (\cpstep{r} \nondet \cestepd) \\
& = (\cpstepd \parallel \cpstep{r}) \nondet (\cpstepd \parallel \cestepd) \nondet 
      (\cestepd \parallel \cpstep{r}) \nondet (\cestepd \parallel \cestepd) \nondet 
      (\cestep{\overline{r}} \Seq \Abort \parallel \cpstep{r}) \nondet (\cestep{\overline{r}} \Seq \Abort \parallel \cestepd) \\
& = \Magic \nondet \cpstepd \nondet 
      \cpstep{r} \nondet \cestepd \nondet 
      \cpstep{(\overline{r} \inter r)} \Seq \Abort  \nondet \cestep{\overline{r}} \Seq \Abort \\
& = \cpstepd \nondet \cestepd \nondet \cestep{\overline{r}} \Seq \Abort .
\end{align*}
Relies distribute over a sequential composition by \refprop{atomic_fp_distrib_seq} 
because a rely command is an iteration of a pseudo-atomic command, and hence it is a pseudo-atomic fixed point.
\begin{align}
  \rely{r} \together c_1 \Seq c_2 & = (\rely{r} \together c_1) \Seq (\rely{r} \together c_2) \labelprop{rely_distrib_seq}
\end{align}
The realisation that pseudo-atomic commands formed an atomic algebra, 
so that we could reuse properties such as \refprop{atomic_sync_iter} and \refprop{atomic_fp_distrib_seq} on pseudo-atomic commands,
led to simpler proofs of \refprop{rely_conj}, \refprop{rely_par_guar} and \refprop{rely_distrib_seq}.
This is an instance where recognising the shared algebraic structure has simplified reasoning.

The proof of law \refprop{rely_guar_intro} below makes use of interchange law between weak conjunction and parallel interchange \refax{conj_interchange_par1},
\begin{align}
  (c_1 \parallel c_2) \together (d_1 \parallel d_2) & \refsto (c_1 \together d_1) \parallel (c_2 \together d_2) \labelax{conj_interchange_par1}
\end{align}
which allows a weak conjunction of $c_1 \parallel c_2$ and $d_1 \parallel d_2$ to be implemented by
the parallel combination of $c_1 \together d_1$ and $c_2 \together d_2$.
Amongst other possibilities, it could also be implemented by parallel composition of $c_1 \together d_2$ and $c_2 \together d_1$.

A useful law allows one side of a parallel command to make use of a rely, 
provided the other side guarantees the rely.
\begin{align}
  \rely{r} \together (c_1 \parallel c_2) & \refsto (\rely{r} \together c_1) \parallel (\guar{r} \together c_2) \labelprop{rely_guar_intro}
\end{align}
Its proof makes use of \refprop{rely_par_guar} and the weak conjunction/parallel interchange law \refax{conj_interchange_par1}:
\begin{align*}
  \rely{r} \together (c_1 \parallel c_2) 
= (\rely{r} \parallel \guar{r}) \together (c_1 \parallel c_2) 
\refsto (\rely{r} \together c_1) \parallel (\guar{r} \together c_2) .
\end{align*}
This then leads to a symmetric version in which the guarantee of each side implies the rely of the other side,
and the original rely, $r$, implies the relies of both sides.
\begin{lemmax}[rely-guar-intro-symmetric]
If  $r \subseteq r_1$ and $r \subseteq r_2$, 
\begin{align*}
  \rely{r} \together (c_1 \parallel c_2) & \refsto (\rely{r_1} \together \guar{r_2} \together c_1) \parallel (\guar{r_1} \together \rely{r_2} \together c_2)
\end{align*}
\end{lemmax}

\begin{proof}
\begin{align*}&
  \rely{r} \together (c_1 \parallel c_2) 
 \Refsto*[as $\together$ is idempotent; weaken relies \refprop{rely_weaken} using assumptions $r \subseteq r_1$ and $r \subseteq r_2$]
  \rely{r_1} \together \rely{r_2} \together (c_1 \parallel c_2) 
 \Refsto*[by \refprop{rely_guar_intro} and parallel is commutative]
  \rely{r_1} \together ((\guar{r_2} \together c_1) \parallel (\rely{r_2} \together c_2)) 
 \Refsto*[by \refprop{rely_guar_intro}]
  (\rely{r_1} \together \guar{r_2} \together c_1) \parallel (\guar{r_1} \together \rely{r_2} \together c_2)
 \qedhere
\end{align*}
\end{proof}

\subsection{Specification commands}\labelprop{specs}

A simple form of specification command, $\Post{t}$, takes a test $t$ and guarantees to terminate in a state satisfying $t$.
It is defined in terms of the command $\Term$ \refdef{term}, which terminates provided it is not preempted by its environment,
(i.e.\ it is scheduled with minimum fairness).
\begin{align}
  \Post{t} & \defs \Term \Seq t  \labeldef{spec}
\end{align}

\begin{lemmax}[parallel-spec]
$\Post{t_1 \meet t_2} = \Post{t_1} \parallel \Post{t_2}$
\end{lemmax}

\begin{proof}
The proof unfolds the specification command \refdef{spec},
applies \reflem{term-par-term} and \refax{test_odot_test},
then \refprop{final_test} on each branch of the parallel,
before folding the result \refdef{spec}:
$
  \Post{t_1 \meet t_2}
 =
  \Term \Seq (t_1 \meet t_2)
 =
  (\Term \parallel \Term) \Seq t_1 \Seq t_2
 =
  \Term \Seq t_1 \parallel \Term \Seq t_2
 =
  \Post{t_1} \parallel \Post{t_2}
$.
\end{proof}
We may now combine \reflem{parallel-spec} with \reflem{rely-guar-intro-symmetric} 
to give a version of the standard rely/guarantee parallel introduction law.
\begin{lemmax}[introduce-parallel]
If $r \subseteq r_1$ and $r \subseteq r_2$,
\begin{align*}
  \rely{r} \together \Post{t_1 \meet t_2} \refsto (\rely{r_1} \together \guar{r_2} \together \Post{t_1}) \parallel (\guar{r_1} \together \rely{r_2} \together \Post{t_2}) .
\end{align*}
\end{lemmax}

\begin{proof}
The lemma follows by applying \reflem{parallel-spec} then \reflem{rely-guar-intro-symmetric} as $r \subseteq r_1$ and $r \subseteq r_2$:
$
  \rely{r} \together \Post{t_1 \meet t_2} 
 =
  \rely{r} \together (\Post{t_1} \parallel \Post{t_2}) 
 \refsto
  (\rely{r_1} \together \guar{r_2} \together \Post{t_1}) \parallel (\guar{r_1} \together \rely{r_2} \together \Post{t_2})
$.
\end{proof}

\section{Conclusions}\labelsect{conclusions}

Restructuring the Concurrent Refinement Algebra (CRA) has taken advantage of 
the common quantale structure of sequential composition, parallel composition, and weak conjunction,
as well as the common biquantale structure that combines pairs of these operators 
and provides, in particular, a weak interchange law for each pair.
Both quantales \cite{2018StruthQuantales} and biquantales are known structures in the literature.
Our earlier work separated out atomic commands \cite{FM2016atomicSteps,FMJournalAtomicSteps}
and recognised that parallel and weak conjunction were both synchronous operations with similar properties.
In our restructure, we have introduced compatible sets of commands,
which are used for both tests and atomic commands,
and treated atomic commands more abstractly.
That allowed us to reuse the atomic algebra theory for pseudo-atomic commands,
and hence get all the laws of atomic commands for pseudo-atomic commands,
simplifying the proofs of a number of properties involving pseudo-atomic comamnds.

\paragraph{Related work.}

In the atomic algebra parallel composition and weak conjunction are both synchronous operations,
in that they synchronise one transition at a time.
Milner's Synchronous Calculus of Communicating Systems (SCCS) \cite{Milner83} introduced a synchronous parallel operator.
Rather than using sequential composition, 
Milner made use of a prefixing operator, $\Ata.c$, 
where the left operand is an element of a set of events (similar to atomic commands in our context) 
and the right operand is a process (similar to a command in our context).
Parallel composition then combines the initial events of two processes, before combine their continuations,
similar to \refax{sync_interchange_odot}.
Using sequential composition, rather than prefixing, allows us to make use of the interchange laws
but we note that sequential composition can be defined in terms of prefixing.

Concurrent Kleene Algebra \cite{DBLP:journals/jlp/HoareMSW11} is the closest comparison to CRA,
especially in the base theories.
The main differences are that:
(a) CKA only treats partial correctness and hence cannot reason about termination,
whereas CRA includes infinite behaviours;
(b)
CKA does not include the abort command, used in CRA for encoding preconditions and rely conditions,
and hence its encoding of satisfying a rely/guarantee specification is stricter than the original of Jones,
in that it requires a thread to satisfy a guarantee even after its rely condition has been violated;
and
(c)
CKA does not include atomic commands 
and hence its distributive laws are only refinements, rather than equalities.

Prisacariu introduced a Synchronous Kleene Algebra (SKA) \cite{Pris10}
similar to Concurrent Kleene Algebra, but which makes use of a synchronous operator,
similar to that used here.
The focus of his work is more on its relationship to automata theory and other theoretical issues, like completeness,
whereas our focus is on developing laws for reasoning about concurrent programs.

\bibliographystyle{alpha}
\bibliography{ms}

\end{document}